\documentclass[slac_one]{revtex4}
\usepackage{graphicx}
\usepackage{fancyhdr}
\usepackage{bm}
\begin{document}

\title{Top $\bm{A_{FB}}$ and charge asymmetry in flavor-dependent chiral
U(1)$^\prime$ model} 

%

\author{P. Ko, Yuji Omura, Chaehyun Yu}
\affiliation{School of Physics, KIAS, Seoul 130-722, Korea}
%

\begin{abstract}
We study the flavor-dependent chiral U(1)$^\prime$ model where only
the right-handed up-type quarks are charged under U(1)$^\prime$ and
additional Higgs doublets charged with U(1)$^\prime$ charges are 
introduced to give proper Yukawa interactions.
We find that in some parameter regions, this model
is not accommodated only with the top forward-backward asymmetry
at the Tevatron, but also with the charge asymmetry at the LHC.
At the same time, the cross section for the same-sign top-quark pair production
in this model could be below the upper limit at the LHC.
\end{abstract}

\maketitle

\thispagestyle{fancy}


\section{Introduction} 
The top forward-backward asymmetry ($A_\textrm{FB}^t$) is one of the most
interesting observables because there exists discrepancy
between theoretical predictions in the standard model (SM) 
and experimental results at the Tevatron. The most recent measurement
for $A_\textrm{FB}^t$ at CDF is 
$A_\textrm{FB}^t=0.162\pm 0.047$ in the letpon+jets channel with a full
set of data~\cite{cdfnew}, which is consistent with
the previous measurements at CDF and D0 within uncertainties~\cite{oldafb}. 
On the other hand,
the SM predictions are between $0.06$ and $0.09$~\cite{smafbac,smafb}.
Since the Tevatron collider shut down last year, this discrepancy will
remain and the origin of the discrepancy might be unresolved.

If the discrepancy in $A_\textrm{FB}^t$ is generated by new physics,
the new physics model would be tested at the LHC. 
Because the LHC is a symmetric $p p$ collider, $A_\textrm{FB}^t$ cannot
be defined, but the charge asymmetry $A_C^y$ defined by
the difference of numbers of events with the positive and negative 
$\Delta |y|=|y_t|-|y_{\bar{t}}|$ divided by their sum, could provide
a test ground for nature of the parity violation in the top quark sector.
The current values for $A_C^y$ are
$A_C^y=-0.018\pm 0.028\pm 0.023$ at ATLAS~\cite{atlasacy} and 
$A_C^y=0.004\pm 0.010\pm 0.012$ at CMS~\cite{cmsacy}, respectively, which are
consistent with the SM prediction $\sim 0.01$~\cite{smafbac}. 
Another interesting observable at the LHC is the cross section for
the same-sign top-quark pair 
production, $\sigma^{tt}$, which is not allowed in the SM. 
The current upper bound on $\sigma^{tt}$ is about 17 pb at CMS~\cite{cmssame}
and 2 pb or 4 pb at ATLAS depending on the model~\cite{atlassame}.
Some models which were proposed 
to account for $A_\textrm{FB}^t$ at the Tevatron,
predict large $A_C^y$ and/or $\sigma^{tt}$ so that they are already disfavored
by present experiments at the LHC.

In this work, we examine the $\textit{so-called}$ 
chiral U(1)$^\prime$ model with flavored Higgs doublets
and flavor-dependent U(1)$^\prime$ charge assignment~\cite{u1models}.
This model is an extension of a $Z^\prime$ model with off-flavor-diagonal
interactions~\cite{zprime}, 
which is excluded by $A_C^y$ and $\sigma^{tt}$ at the LHC.
In the Refs.~\cite{u1models}, the authors 
proposed a model with chiral U(1)$^\prime$ symmetry for the construction
of a realistic $Z^\prime$ model with flavor-off-diagonal couplings,
where only the right-handed up-type quarks are charged under U(1)$^\prime$,
and the $Z^\prime$ boson is identified as a gauge boson of the U(1)$^\prime$
symmetry. Then, in order to construct a realistic Yukawa interactions,
additional Higgs doublets with U(1)$^\prime$ charges should be introduced.
This is quite similar to the unitarity problem of 
the $W_L W_L \to W_L W_L$ or $f\bar{f} \to W_L W_L$ scatterings 
in the intermediate vector boson model, where the unitarity is recovered
by introducing a Higgs doublet. This is also true in any other models
with a new spin-1 particle and chiral U(1)$^\prime$ charges. 
Finally, new chiral fermions have to be introduced in order to cancel
the gauge anomaly. For more details of the chiral U(1)$^\prime$ models,
we refer readers to Refs.~\cite{u1models}. We point out that the simple 
$Z^\prime$ model may be disfavored by the experiments at the LHC, but
if one considers more complete model, then the extended model could be
revived.

In this proceeding, we consider two scenarios of the chiral U(1)$^\prime$
model. One is a light $Z^\prime$ boson case 
with a neutral scalar Higgs boson $H$
and a pseudoscalar Higgs boson $a$. The other is a light Higgs boson $h$ case
with a heavier $Z^\prime$ boson, a heavier Higgs boson $H$, and 
a pseudoscalar Higgs boson $a$, motivated by the recent observation
of an SM-like Higgs boson at the LHC~\cite{higgs}. 
The other particles newly
introduced in the model are assumed to be heavy or have small couplings
so that they are decoupled from top physics. 
Finally, we present a summary of this work.

\begin{figure*}[t]
\centering
\includegraphics[width=110mm]{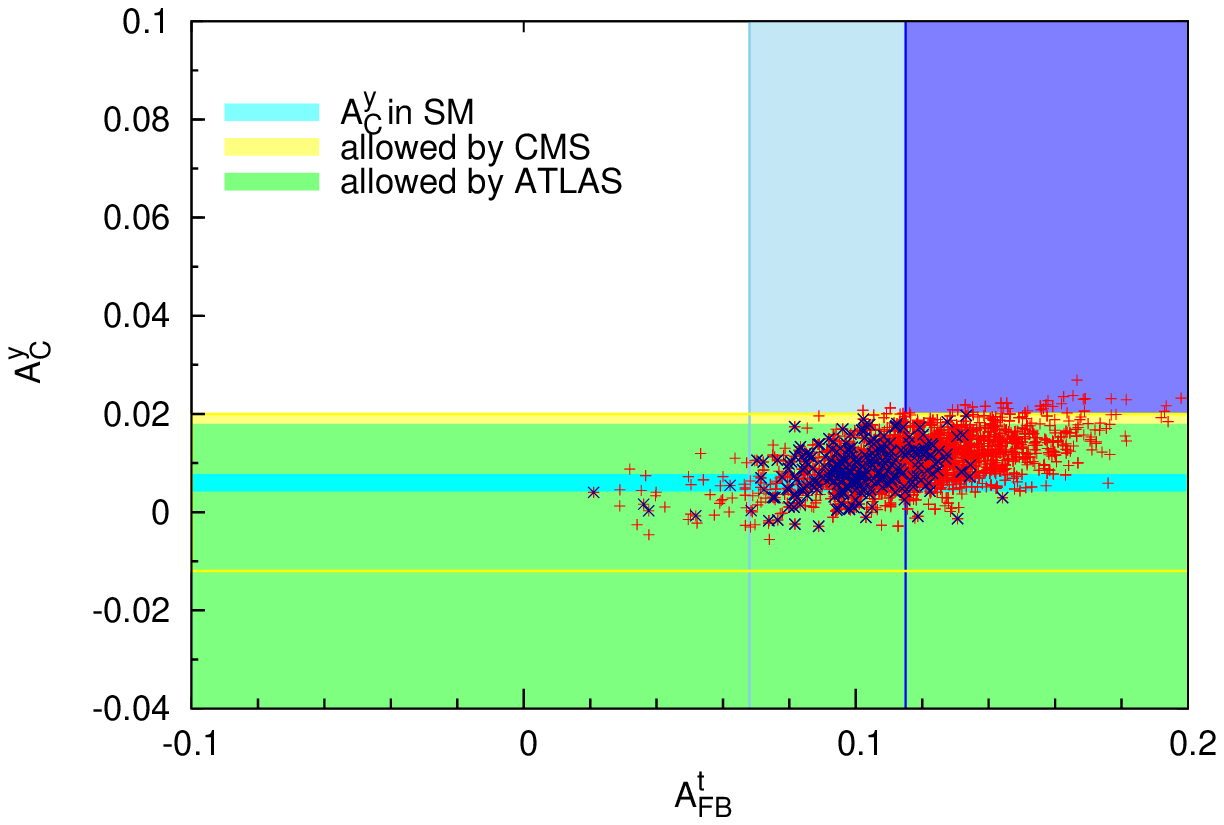}
\caption{$A_\textrm{FB}^t$ at the Tevatron and $A_C^y$ at the LHC
for $m_{Z^\prime}=145$ GeV.} 
\label{fig1}
\end{figure*}

\section{Light $\bm{Z^\prime}$ boson case}
In this section, we consider a light $Z^\prime$ boson case with a mass
$m_{Z^\prime}=145$ GeV. In order to suppress the non-SM decay of the top quark,
we require that the Higgs bosons $H$ and $a$ are heavier than the top quark.
However, this requirement might be inconsistent with the recent observation
of an SM-like Higgs boson at the LHC~\cite{higgs} 
and with non-observation of a Higgs-like signal in a large region 
between 130 GeV and 600 GeV~\cite{nonhiggs}. 
In order to accommodate these results, we assume that the $u$-$t$-$h$ coupling,
where $h$ is the lightest Higgs boson,
is small enough to be decoupled from the top-quark pair production.
In this model, the $Z^\prime$ boson can contribute to the top-quark pair
production through its $s$-channel and $t$-channel exchanges
in the $u\bar{u}\to t\bar{t}$ process. While the Higgs bosons contribute
to the top-quark pair production
only in the $t$ channel because their Yukawa couplings to light quarks
are negligible.
We scan the following parameter regions:
$180~\textrm{GeV} \le m_{H,a} \le 1~\textrm{TeV}$,
$0.005 \le \alpha_x \le 0.012$,
$0.5 \le Y_{tu}^{H,a} \le 1.5$, and
$(g_R^u)_{tu}^2=(g_R^u)_{uu} (g_R^u)_{tt}$, 
where $\alpha_x$ is a U(1)$^\prime$ gauge
coupling and $Y_{tu}^{H,a}$ are flavor-off-diagonal 
Yukawa couplings. One can also consider the case where the $s$-channel exchange
of the $Z^\prime$ boson is negligible by setting the coupling $(g_R^u)_{uu}=0$,
but the numerical results are not so different.

In Fig.~\ref{fig1}, we show the scattered plot for $A_\textrm{FB}^t$ at the
Tevatron and $A_C^y$ at the LHC.
The green and yellow regions are consistent with $A_C^y$ at ATLAS and CMS
in the $1\sigma$ level, respectively. The blue and skyblue regions are
consistent with $A_{FB}^t$ in the lepton+jets channel at CDF in the
$1\sigma$ and $2\sigma$ levels, respectively.
The red points are in agreement with the cross section 
for the top-quark pair production
at the Tevatron in the $1\sigma$ level and the blue points are consistent with
both the cross section for the top-quark pair production at the Tevatron
in the $1\sigma$ level and the upper bound on the same-sign top-quark pair
production at ATLAS. We find that a lot of parameter points can explain all
the experiments: the total cross section, the forward-backward asymmetry,
the same-sign top-pair production, and the top charge asymmetry, which are
considered in this work. We emphasize that the simple $Z^\prime$ model
is excluded by the same-sign top-quark pair production, but
in the chiral U(1)$^\prime$ model, this strong bound could be evaded
due to the destructive interference between the $Z^\prime$ boson and Higgs
bosons.

\begin{figure*}[t]
\centering
\includegraphics[width=110mm]{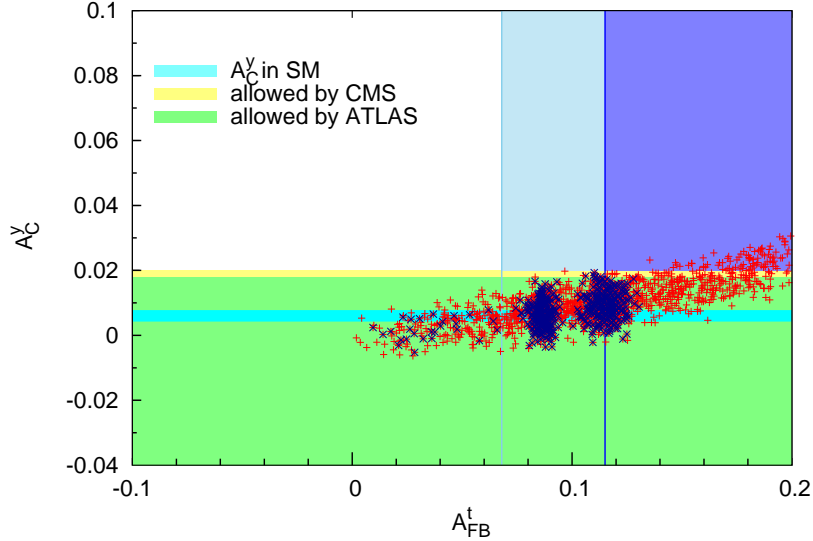}
\caption{$A_\textrm{FB}^t$ at the Tevatron and $A_C^y$ at the LHC
for $m_h=125$ GeV.} 
\label{fig2}
\end{figure*}

\section{Light Higgs boson case}
In this section, we discuss a light Higgs boson case with $m_h=125$ GeV,
motivated by the recent observation of an SM-Higgs like scalar boson 
at the LHC~\cite{higgs}. In this case, the $Z^\prime$ boson and Higgs bosons
$h$, $H$, and $a$ contribute to the top-quark pair production.
In order to suppress the exotic decay of the top quark into $h$ and $u$, 
we set the Yukawa
coupling of $h$ to be $Y_{tu}^h \le 0.5$ and masses of $Z^\prime$, $H$,
and $a$ are larger than the top-quark mass or approximately equal to the
top-quark mass. We scan the following parameter regions:
$160~\textrm{GeV} \le m_{Z^\prime} \le 300~\textrm{GeV}$,
$180~\textrm{GeV} \le m_{H,a} \le 1~\textrm{TeV}$,
$0 \le \alpha_x \le 0.025$,
$0 \le Y_{tu}^{H,a} \le 1.5$ and 
$(g_R^u)_{tu}^2=(g_R^u)_{uu} (g_R^u)_{tt}$. 
The mass region of the $Z^\prime$ boson
is taken to avoid the constraint from the $t\bar{t}$ invariant mass
distribution at the LHC. If $(g_R^u)_{uu}\simeq0$ 
and the $s$-channel contribution of the $Z^\prime$ could be ignored,
the mass region of the $Z^\prime$ boson could be enlarged.

In Fig.~\ref{fig2}, we show the scattered plot for $A_\textrm{FB}^t$ 
at the Tevatron and $A_C^y$ at the LHC for $m_h=125$ GeV.
All the legends on the figure are the same as those in Fig.~\ref{fig1}.
We find that there exist parameter regions which agree with all the experimental
constraints considered in this work. We emphasize that in some parameter spaces
$\sigma^{tt}$ is less than 1 pb.

\section{Summary}
The top forward-backward asymmetry at the Tevatron
is the only quantity which has 
deviation from the SM prediction in the top quark sector up to now.
A lot of new physics models have been introduced to 
account for this deviation
or it has been analyzed in a model-independent way~\cite{modelindep}.
Some models have already been disfavored by experiments at the LHC and
some new observables are introduced to test the models~\cite{modelindep}.
In this work, we investigated the chiral U(1)$^\prime$ model with flavored
Higgs doublets and flavor-dependent couplings. Among possible scenarios,
we focused on two scenarios: a light $Z^\prime$ boson case and 
a light Higgs boson case. We found that both scenarios can be accommodated with
the constraints from the same-sign top-quark pair production and
the charge asymmetry at the LHC as well as the top forward-backward asymmetry
at the Tevatron. In the light $Z^\prime$ scenario, the same-sign top-quark
pair production at the LHC would exclude the scenario if its upper bound
becomes less than 1 pb. However, in the light Higgs boson scenario,
we could find some parameter regions where the cross section for the same-sign
top-quark pair production at the LHC is less than 1 pb.

The chiral U(1)$^\prime$ model has a lot of new particles except for
the $Z^\prime$ boson and neutral Higgs bosons. The search 
for exotic particles
may constrain our model severely. For example, our model is strongly
constrained by search for the charged Higgs boson in the
$b\to s\gamma$, $B\to \tau \nu$, and $B\to D^{(\ast)}\tau \nu$ decays.
In order to escape from such constraints, we must assume a quite heavy charged
Higgs boson or it is necessary to study our model more carefully
by including all the interactions which have been neglected in this work.
Search for the dijet resonances would also give strong constraints 
on the $Z^\prime$ boson. If the $s$-channel contribution is not negligible,
the coupling $(g_R^u)_{uu}$ is constrained by the search for the dijet 
resonances. However, if the $s$-channel contribution can be ignored, 
we can avoid the constraints from the search for the dijet resonances
as well as from the $t\bar{t}$ invariant mass distribution.
In the chiral U(1)$^\prime$ model, new chiral fermions must be included
for the anomaly cancellation. Then, search for the exotic fermions like
the 4th generation fermions would also constrain our model. Furthermore,
there could be cold dark matter candidates in our model
so that the dark matter experiments
would give constraints. The most severe constraints arise from the search
for the Higgs boson. In this work, 
we discussed the cases where $m_h=125$ GeV, but
we did not take into account its branching ratios.
If the branching ratios of the SM-like Higgs boson observed at the LHC
settle down at the present values, our model will severely be constrained.
We emphasize that this study is not complete yet, because there are extra
fields which are subdominant in the top-quark production.
To a complete study, we need to consider the Higgs phenomenology more carefully.

{\it Note Added}

While we are finalizing this work, the CMS collaboration
has announced much stronger upper limit on the same-sign top-quark
pair production: $\sigma^{tt} \le 0.39$ pb 
at 95\% confidence level~\cite{cmsttnew}.
This bound would exclude the light $Z^\prime$ boson case, but
we note that the light Higgs boson case could still be consistent
in a certain parameter space.

\begin{acknowledgments}
This work is supported in part by Basic Science Research Program
through NRF 2011-0022996 and in part by NRF Research Grant
2012R1A2A1A01006053.
\end{acknowledgments}

\end{document}